\documentclass[%
 reprint,
 amsmath,amssymb,
 aps,
]{revtex4-2}

\usepackage{threeparttable}
\usepackage{upgreek}
\usepackage{graphicx}
\usepackage{dcolumn}
\usepackage{bm}
\usepackage{lipsum}
\usepackage{mathrsfs}
\usepackage{xcolor}

\begin{document}

\title{Energy resolving optical transition-edge sensor approaching fundamental limits}

\author{Leaf Swordy,$^{1}$ Benedikt Hampel,$^{1}$ Paul Szypryt,$^{1,2}$ Negar Otrooshi,$^{1,2}$ Douglas A. Bennett,$^{1}$, Jordan Wheeler,$^{1}$ Galen O'Neil,$^{1}$ Michael R. Vissers,$^{1}$ Daniel S. Swetz,$^{1}$ Joel N. Ullom,$^{1}$  Martin J. Stevens,$^{1}$ Adriana E. Lita$^{1}$}

\address{1. National Institute of Standards and Technology, Boulder, Colorado 80305, USA}

\address{2. Department of Physics, University of Colorado, 2000 Colorado Avenue, Boulder, Colorado 80309, USA \\ leaf.swordy@nist.gov$^*$}


\begin{abstract}
Transition-edge sensors (TES) have long held promise for exhibiting single-photon spectral resolution with near-unity detection efficiency, no read noise, and negligible dark counts. A TES detector capable of simultaneously achieving these traits could be revolutionary in fields including quantum computing, quantum sensing, biological imaging, and the detection of life on alien worlds by spectroscopy of exoplanet atmospheres. The applicability of optical TES detectors in these fields has been limited by energy resolution results that differ significantly from the theoretical expectations. We report significant enhancement of the single-photon energy resolution of tungsten (W) TESs by mitigation of athermal phonon loss during the photon-absorption energy down-conversion cascade. Implementing barriers to phonon loss dramatically increased the TES energy resolution ($\Delta E$). We achieve $\Delta E$ of 34 meV - 61 meV, with resolving power ($R=E/\Delta E$) of 24 - 43, in a band spanning 1550 nm - 471 nm. All relevant electrical/thermal device parameters are explored in order to model theoretical limits to $\Delta E$ imposed by noise, and provide an estimate of the athermal phonon trapping factor. Based on a simple model of electrical, thermal, and phonon escape noise sources, we estimate a phonon trapping factor approaching $90\%$ for our best performing TES architecture. These results bring optical TES technology closer to theoretical predictions and to achieving the ambitious goals of photon-starved optical spectroscopy and fault-tolerant photonic quantum computing.

\end{abstract}

\keywords{Transition Edge Sensor, TES, Superconducting Detectors, Energy Resolution, Astronomy, Phonon Trapping, Habitable Worlds}
\maketitle

\section{Introduction}
 Transition-Edge Sensors (TESs) \cite{irwin2005} are superconducting devices with long-demonstrated ability as both bolometric (total irradiance), and calorimetric (single photon) photodetectors. TES-based detectors have been widely used to detect photons over a broad spectrum with applications ranging from millimeter-wave astrophysics \cite{Hubmayr2018} to quantum optics \cite{Gerrits2016} and x-ray and gamma-ray spectroscopy \cite{Bennett}\cite{Doriese2017}. In the VIS-NIR regime, the optimization of TES detectors for the past decade focused on quantum information  (QI) applications that require near-unity single-photon detection efficiencies and fast count rates. Recent years have seen advancements across many optical TES characteristics including energy resolution \cite{Hattori2022}, quantum efficiency \cite{Larsen2025}, and recovery time \cite{2023JAP...133w4502H}. Many of these advancements have been pursued in the context of QI applications such as quantum computing \cite{xan} and quantum  networking \cite{Jabir:24}. In these applications, the ability of a TES to estimate the energy of individual photons via pulse analysis is vital for distinguishing between single photon absorption events, and events consisting of more than one photon ($N = 1, 2, 3, 4, ...$). 


Moreover, the ability of TES detectors to resolve single photons, while simultaneously giving an estimate of the absorbed photon energy, has potential application in fields ranging from single-photon microscopy \cite{Fukuda2018}, and dark matter searches \cite{Schwemmbauer} to spectroscopy of exoplanet atmospheres \cite{Nagler2021}. Applicability of TESs to these science cases has been hindered by energy resolution results that differ significantly from the theoretically predicted resolution floor \cite{Wessels2021,Hattori2022}. Causes of this discrepancy include unexplained `excess' noise \cite{Hattori2022}, and loss of athermal phonons early in the energy down-conversion cascade \cite{Kozorezov2006}. In this work, we aim to mitigate the degradation of energy resolution resulting from athermal phonon loss by suspending tungsten (W) TESs on thin silicon nitride (SiN$_\textrm{x}$) membranes. Detailed noise analyses are pursued in order to understand the differences between the measured energy resolution and the theoretically predicted resolution.


Our optical TES devices, previously optimized for QI applications, are fabricated from thin W films patterned in a square area of 20 $\upmu$m x 20 $\upmu$m 
and have superconducting transition temperatures ($T_{\textrm{c}}$) of $\sim$180 mK. When fabricated on a silicon substrate, these devices typically achieve $R=E/\Delta E \approx 4$ resolving power at 1550 nm wavelength (200 meV full width at half maximum (FWHM) energy resolution at 800 meV photon energy) \cite{Gerrits2016}. The $T_{\textrm{c}}$ is a device parameter that directly affects the device energy resolution, which scales as $T_{\textrm{c}}^{3/2}$. Hence, a lower $T_{\textrm{c}}$ will determine an improved device energy resolution \cite{irwin2005}. 

Previous attempts to enhance energy resolution of our W TESs by suspension on SiN$_\textrm{x}$ membranes (500 nm thickness) showed improvements to energy resolution from 150 meV to 90 meV for devices with $T_{\textrm{c}}$ $\approx{ 160}$ mK\cite{ltd_2026}. 
In this work we explore further improvements in energy resolution by both device design, including lowering $T_{\textrm{c}}$, as well as optimization of membrane thickness for enhanced phonon trapping for our device and material parameters. A simple phonon ray-tracing simulation adapted from de Visser et al. \cite{PhysRevApplied.16.034051} to our detector materials and geometry, demonstrated (in a heuristic fashion) very aggressive scaling of the phonon ``dwell-time'' (the total time the phonon spends within the superconducting detector film before escaping to the substrate) with decreasing membrane thickness. This preliminary study indicated potential benefits for fabrication on extremely thin membranes ($<$100 nm). The TES devices studied in this work are fabricated on 75 nm thick SiN$_\textrm{x}$ membranes and
are characterized in terms of their device parameters, as detailed in the supplementary material, including the thermal/electrical 
sensitivities ($  \alpha =  \frac{T_0}{R_0}\frac{\partial R}{\partial T} |_{I_0}, \; \; 
    \beta = \frac{I_0}{R_0}\frac{\partial R}{\partial I} |_{T_0}$),  thermal conduction to the substrate ($k = G/nT^{n-1}$, where $G$ is the TES thermal conductance and $n$ is the temperature exponent of the power flow into the substrate), TES heat capacity ($C$), and their noise characteristics.


    Measurements of TES device energy resolution ($\Delta E_{\textrm{meas}}$) were obtained from device response to an attenuated pulsed laser.  
    The measured energy resolution was then compared with theoretical expectations from models of energy resolution and noise in order to better understand the effects of athermal phonon escape. Ultimately, we find that athermal phonon loss is a very significant factor in limiting our device resolution compared to theoretical predictions that do not account for athermal phonon loss. 

    Using pulsed laser light at 1550 nm, 1064 nm, 800 nm, and 471 nm we are able to quantify the enhanced resolution, and find that increased trapping brings our measured resolution closer to that predicted by models without athermal phonon escape. The results for our best performing device design 
    represent a new energy resolution record for TESs operating in the VIS-NIR range, with 34 meV FWHM  at 0.8 eV (1550 nm).

\section{Optical TES Device Design}

\begin{figure}
\centering\includegraphics[width=3.2in]{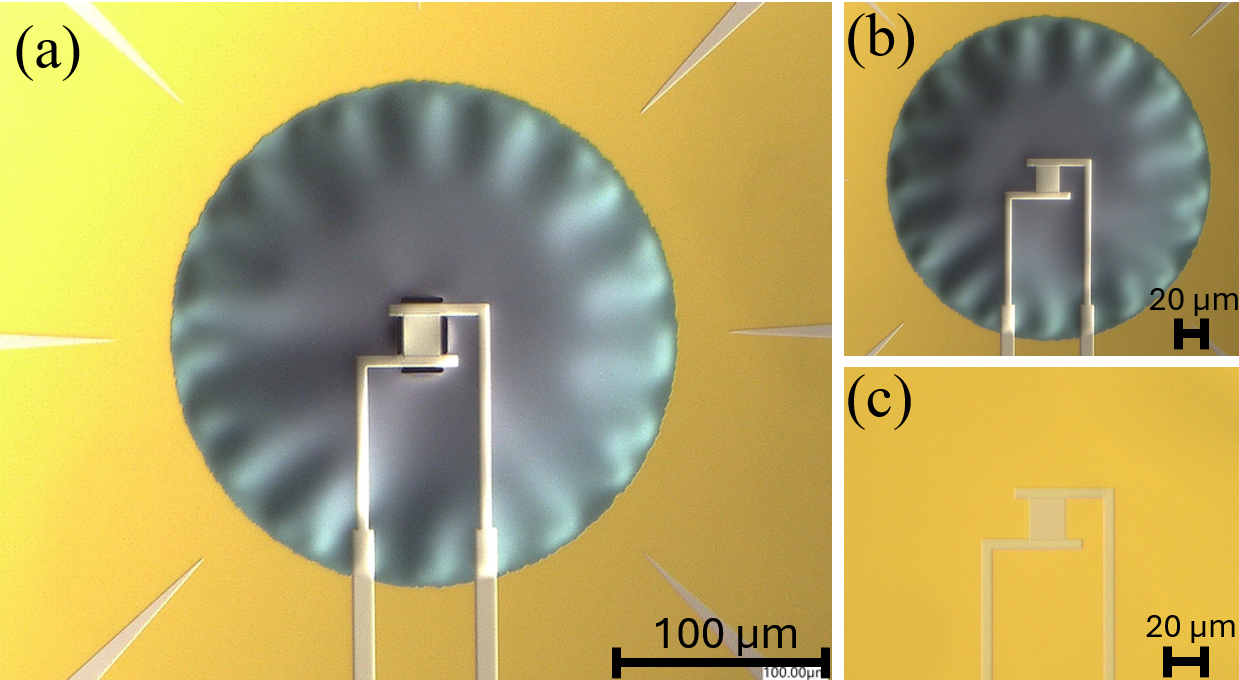}
\caption{\label{fig:epsart} Three designs of low $T_{\textrm{c}}$ TES to investigate the effects of athermal phonon loss on TES energy resolution. (a) TES suspended on a 75 nm SiN$_\textrm{x}$ membrane with micro-cutouts directly adjacent to the W device, dubbed `TES MC'. This device scheme yielded the best resolution results, as will be discussed in Section \ref{sec:eres}. (b) TES suspended on a 75 nm SiN$_\textrm{x}$ membrane without micro-cutouts (TES M). (c) TES device on a 75 nm SiN$_\textrm{x}$ layer on top of the silicon substrate, with a 120 nm SiO$_2$ etch-stop layer in between (TES S). The photosensitive W device area is 20 $\upmu$m $\times$ 18 $\upmu$m for all three schemes.} 
\end{figure}

Our tungsten-based TES detectors consist of a W trilayer with 22 nm  W film and a $\sim4$ nm amorphous Si layer underneath and above the W film, biased with 80 nm niobium wire leads (Fig \ref{fig:epsart}). The aSi/W/aSi trilayer design ensures a fixed ratio of the alpha and beta crystallographic phases of tungsten and also protects the device from oxidation. The W trilayer film was patterned using electron-beam lithography for smooth edges \cite{ltd_2026} and wet etched such that a 20 $\upmu$m $\times$ 18 $\upmu$m  rectangle is exposed to photons and 2 $\upmu$m $\times$ 18 $\upmu$m is under each of the Nb leads. Our previous studies have shown that a smoother TES edge resulting from electron-beam lithography compared to optical lithography, resulted in higher device $T_{\textrm{c}}$ and critical current. The device substrate consisted of a Si wafer with thermally grown 75 nm SiN$_\textrm{x}$/ 120 nm SiO$_2$ films. The SiO$_2$ layer provided an etch stop during the removal of Si for SiN$_\textrm{x}$ membrane fabrication, and no efforts were done to remove any residual SiO$_2$. 

We used three designs to investigate the effects of athermal phonon loss on TES energy resolution, as shown in Fig. \ref{fig:epsart}. Using a deep reactive ion etching (RIE) process the Si  was removed from an area of 250 $\upmu$m  diameter, allowing the TES devices to be suspended on the SiN$_\textrm{x}$ membranes (Fig \ref{fig:epsart}a-b), labeled as TES MC and TES M, respectively; while other TES devices (Fig \ref{fig:epsart}c) were left on the solid SiN$_\textrm{x}$/SiO$_2$/Si wafer, identified as TES S. Some of the devices on membranes (Fig \ref{fig:epsart}a), TES MC, also had micro-cutouts next to the TES to prevent lateral phonon escape. The same deep RIE process used to release the membranes was used to etch through the Si wafer substrate and release the detector dies in a keyhole pattern, ready for mounting and wirebonding. The keyhole pattern enables the use of a self-alignment technique that ensures  less than 1 \% coupling loss \cite{Miller:11} between the optical fiber tip and TES detector. This alignment scheme prevents energy resolution degradation due to absorption in the Nb wiring instead of detector area.
A single mode 1064 nm optical fiber with a 5.3 $\upmu$m core diameter was coupled to the TES detector.

The TES was connected in parallel with a 4.5 m$\Omega$ shunt resistor for voltage biasing and was read out using a two-stage Superconducting Quantum Interference Device (SQUID) readout. The TES and the two-stage SQUID readout are co-packaged to minimize parasitic inductance and provide lower input inductance readout\cite{2023JAP...133w4502H}, and they are mounted on the 50 mK stage of an adiabatic demagnetization refrigerator (ADR). A superconducting shield was used in order to prevent TES performance degradation due to magnetic field effects \cite{ltd_2026}.


\section{Measurements and Modeling}

\begin{table}
\centering\caption{\label{tab:params}
Summary of relevant TES device parameters as determined from measurements of pulse data, IV curves, noise spectra, complex impedance, and heat capacity. All values are determined using the TES model outlined in the supplementary material.}
\begin{threeparttable}
\begin{tabular}{lccccc}
 &unit& TES MC$^\textrm{a}$  & TES M$^\textrm{b}$  & TES S$^\textrm{c}$  &
\\
\hline
$T_\textrm{c}$ & mK& 114 & 109 & 105
\\
$R_n$ & $\Omega$& 5.37 & 5.46 & 5.75 
\\
$n$ && 4.16 & 4.04 & 4.8 
\\
$G$& pW/K& 1.58  & 2.2  & 1.93 
\\
$C$& fJ/K& 0.183 & 0.204 & 0.219 
\\
$\alpha$&& 237$^\textrm{d}$ & 200$^\textrm{d}$ & 205$^\textrm{d}$
\\
$\beta$&& 1.43$^\textrm{d}$ & 1.73$^\textrm{d}$ & 1.18$^\textrm{d}$
\\
$\mathscr{L}$&& 54.2 & 47.3 & 41.4 
\\
$\tau^-_{\textrm{meas}}$ & $\upmu$s & 5.32$^\textrm{e}$ & 6.47$^\textrm{e}$  & 7.58$^\textrm{e}$
\\
$\tau^-_{\textrm{calc}}$ & $\upmu$s & 5.56$^\textrm{f}$ & 6.76$^\textrm{f}$ & 8.55$^\textrm{f}$
\\
$\Delta E_{\textrm{meas}}$& meV & 34$^\textrm{e}$  & 40$^\textrm{e}$  & 54$^\textrm{e} $ 
\\
$\Delta E_{\textrm{noise}}$& meV & 21$^\textrm{g}$ & 28$^\textrm{g}$ & 30$^\textrm{g}$
\\
$\Delta E_{\textrm{full trapping}}$& meV & 19& 21 & 20
\\
\end{tabular}
\begin{tablenotes}
\scriptsize
\item[a]{75 nm SiN$_\textrm{x}$ membrane TES with micro-cutouts, Fig. \ref{fig:epsart}a.}
\item[b]{75 nm SiN$_\textrm{x}$ membrane TES without micro-cutouts, Fig. \ref{fig:epsart}b.}
\item[c]{TES on 75nm SiN$_\textrm{x}$ / 120 nm SiO$_2$ / Si substrate, membrane not released,
Fig. \ref{fig:epsart}c.}
\item[d]{At the point in transition yielding the highest $E/\Delta E$. 13\% $R_\textrm{n}$, 11\% $R_\textrm{n}$, and 20\% $R_\textrm{n}$ respectively.}
\item[e]{Measured from 1550 nm pulsed laser source.}
\item[f]{Estimated from complex impedance measurements.}
\item[g]{Measured noise and TES responsivity from complex impedance.}
\end{tablenotes}
\end{threeparttable}
\end{table}

\subsection{\label{sec:noise} Thermodynamic Noise} 

\begin{figure}
\centering\includegraphics[width=3.0in]{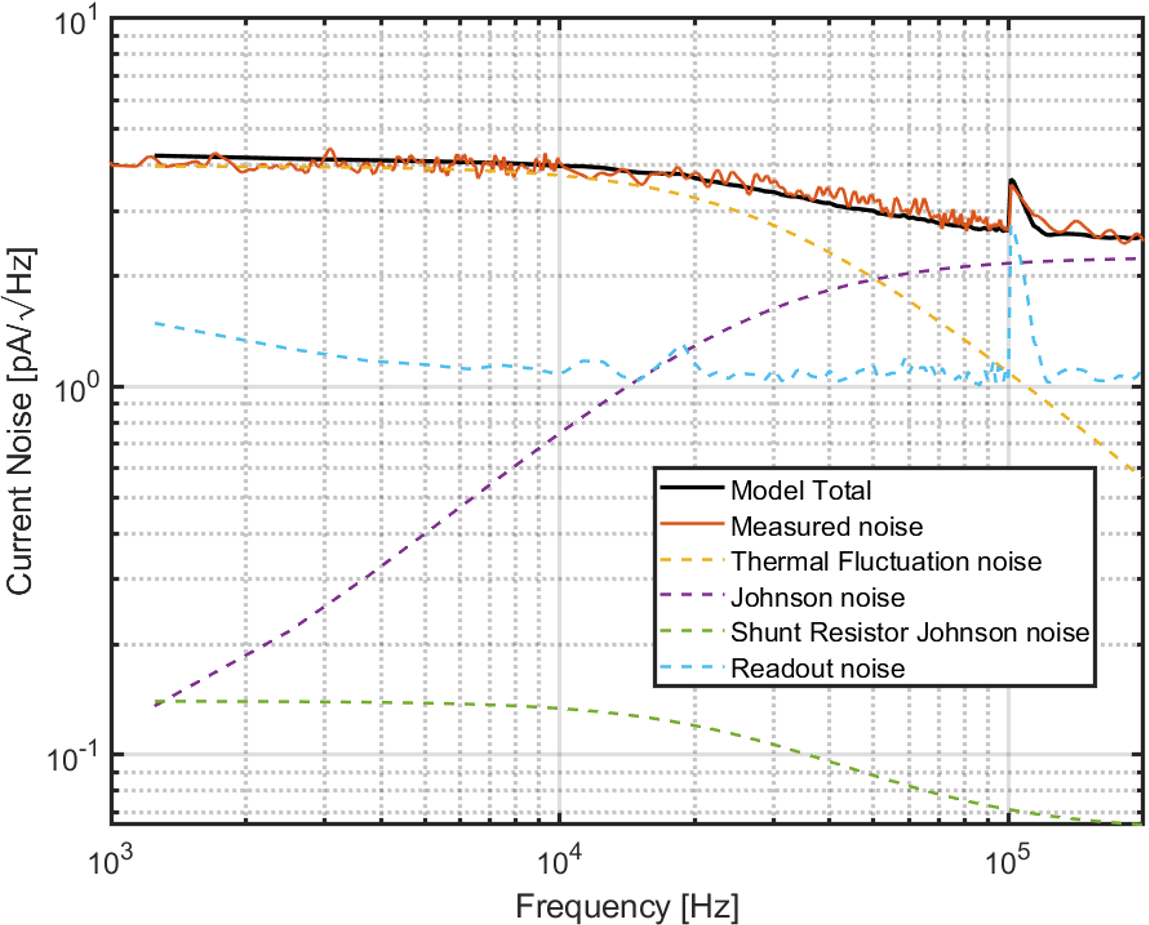}
\caption{\label{fig:noise} Theoretical and measured current noise for the TES MC device biased at the point in transition yielding the highest $E/\Delta E$ resolution ($13\% \; R_n$). The overall theoretical noise, the sum of the electrical noise (purple and green curves), thermal noise (orange curve), and experimentally measured readout noise (blue curve), is shown as the solid black line, calculated using the device parameters of Table \ref{tab:params} with the device model outlined in the supplemental material. This prediction is seen to be in good agreement with the TES experimentally measured noise (red curve), computed as the average of 200 noise traces.} 
\end{figure}

TES detectors are susceptible to thermodynamic noise due to fluctuations in both current (Johnson noise) and temperature (thermal fluctuation noise). To investigate the difference between the theoretical expectation and the measured value, the current noise of the detector was measured
and compared with the theoretical model. Figure \ref{fig:noise} shows the noise components calculated using the single-block thermal model \cite{irwin2005} as described in the supplemental material. 
The theoretical current noise labeled as `Model Total' in  Fig. \ref{fig:noise} comprises contributions from thermal fluctuation noise (TFN), Johnson noise of TES and of the shunt resistor,
and readout amplifier noise. We modeled each contribution using the TES small-signal resposivity ($s_I(\omega)$) estimated from
 extracted parameters from complex impedance measurements  (as shown in supplemental material). Table \ref{tab:params} lists the device parameters extracted from complex impedance measurements such as $\alpha$ and $\beta$ 
 as well as other device parameters used for noise components estimation such as loop gain $\mathscr{L}$ defined as $\mathscr{L} = I_0^2R_0\alpha/GT_0$, (where $R_0$ and $I_0$ and $T_0$ are the TES resistance, current and temperature at operating point), thermal conductance $G$ and heat capacity $C$.
The thermal conductance between the TES and the heat bath is an input parameter for the TFN noise. The Joule power dissipation in the voltage-biased TES is equal to the heat flow to the heat bath. 
By measuring the Joule power dissipation at different bath temperatures, we extracted the thermal conductance $G$  at $T_c$. 
The TES Joule heating power vs bath temperature data was fitted by using a power law for thermal conductance ($G = nkT^{n-1}$), while the $T_c$, $n$ and $k$ (thermal conductivity)
are left as free parameters. The measured values for the thermal conductances, $G$, varied from 1.58 to 2.2 pW/K, as shown in Table \ref{tab:params}.  The thermal exponent $n$ (4.8, 4,16 and 4.04) is determined by the nature of the thermal link to the heat bath. A value of $n \approx 5$ indicates that the dominant thermal bottleneck is the weak coupling between electrons and phonons within the superconducting film itself (standard $T^5$ power law for low-temperature electron-phonon scattering). Whereas a value of $n \approx 4$ indicates that the primary thermal impedance is the thermal boundary resistance (Kapitza resistance) between the TES film and the substrate, or 2D acoustic phonon transport through the membrane/substrate to the bath. The values of $n$ attained for the three device designs demonstrate that, as we increasingly isolate the device from the substrate, we move from an electron-phonon decoupling dominated system, to a Kapitza resistance dominated system. 

\begin{figure*}
\centering\includegraphics[width=4.3in]{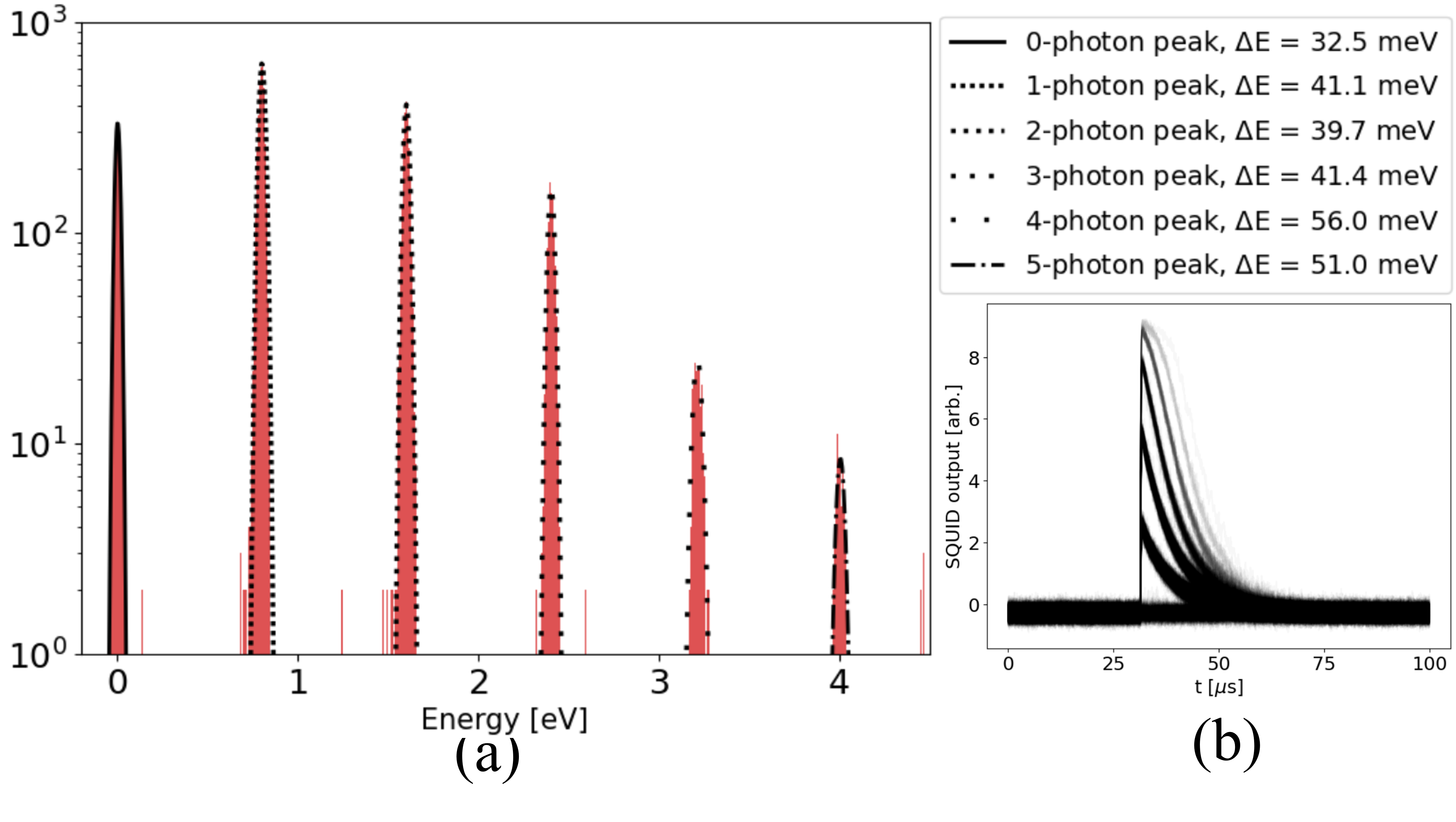}
\caption{\label{fig:15501}1550 nm pulsed laser light incident on the TES device of Fig. \ref{fig:epsart}a. Raw pulse data for 10,000 traces (b) is analyzed via optimal filtering \cite{Szymkowiak1993Signal} and subsequently binned in a histogram. The locations of the 0-5 photon peaks are used to calibrate the raw data pulse-height to meV. The histogram is given in semi-log scale, with fitted Gaussians to extract the FWHM $\Delta E$ energy resolution of each filtered peak. The fraction of pulses at risk of being `miscounted' (i.e. binned into a histogram peak corresponding to the wrong photon number) is less than 1/1000.}
\end{figure*}

The noise plots in Fig. \ref{fig:noise} correspond to the TES MC device biased at the point in transition giving the highest measured $E/\Delta E$ (13$\%\;R_n$, near the point of highest $\alpha$). 
We also plot the measured noise (an average of 200 noise spectra) 
for both the SQUID readout in isolation, when the TES is normal (Readout noise), as well as the entire TES + readout system (Measured noise). We see from Fig. \ref{fig:noise} that the theoretical noise (Model Total, black curve) expressed as the sum of the thermal fluctuation noise, the Johnson noise and the readout noise (as detailed in the supplementary document) is in good agreement with the experimentally measured noise (red curve). Notably, the noise model does not include any unexplained excess Johnson noise term ($M^2 = 0$) \cite{Hattori2022}. Thermal fluctuation noise and Johnson noise were calculated from parameters
obtained via complex impedance fitting, while readout
noise is the measured SQUID readout noise.

In the small-signal limit, we can calculate the expected energy resolution of devices by integrating over the total measured noise spectrum as shown in Eq. 7 \cite{McCammon2005} in Supplemental material. The 
expected energy resolution calculated from the noise measurements, is labeled as $\Delta E_{\textrm{noise}}$ and is shown for all three device designs in Table \ref{tab:params}. This energy resolution value, normalized by the fraction of captured energy, will correspond to the TES measured energy value in the absence of athermal phonon losses. (If all the energy were captured in the electron system with no athermal phonon escape then this noise-limited energy resolution should match the measured energy resolution)

\subsection{\label{sec:eres}Measured Energy Resolution}

TES energy resolution was characterized using highly attenuated pulsed laser light at 1550 nm, 1064 nm, 800 nm, and 471 nm for the TES MC device (shown in Fig. \ref{fig:epsart}a). Fig. \ref{fig:15501}b shows the results of 10,000 pulse traces at 1550 nm, the wavelength for which the device was originally optimized. The raw pulse data of Fig. \ref{fig:15501}b is processed with an optimal filter \cite{Szymkowiak1993Signal}, and subsequently binned in a pulse height histogram (Fig. \ref{fig:15501}a). Gaussian functions are fit to each photon number peak, and the Full Width at Half Maximum (FWHM) is taken as $\Delta E_{\textrm{meas}}$.

\begin{figure*}
\centering\includegraphics[width=3.0in]{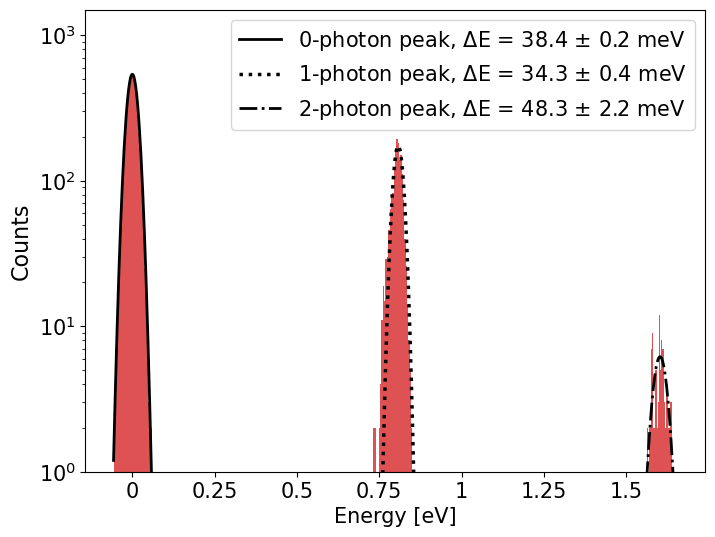}
\caption{\label{fig:15502}1550 nm pulse data collected in the same manner as Fig. \ref{fig:15501} with a much lower photon flux, and with the device of Fig. \ref{fig:epsart}a biased at the optimum conditions for separating the 0 photon, 1 photon, and 2 photon peaks ($\sim13\% \; R_n$).}
\end{figure*}

Figure \ref{fig:15501}a demonstrates that our low $T_\textrm{c}$ optical TES can absorb $\sim5$ photons at 1550 nm before saturation. Fig. \ref{fig:15502} shows the results of a highly attenuated source, biased at the point giving the best resolution of $\Delta E = 34 \pm0.4$ meV for the single-photon peak at 1550 nm. This analysis was also performed at 1550 nm for the TES M and TES S devices shown in Fig. \ref{fig:epsart}b-c, giving a resolution of $40\pm0.7$ meV and $54\pm0.8$ meV respectively. The resolution difference between Figs. \ref{fig:15501} \& \ref{fig:15502} demonstrates that if the detector is biased low in the transition to achieve the full dynamic range of the device, we degrade the resolution by $\sim6$ meV compared to when we optimize only for the 1-photon peak. Fig. \ref{fig:15501} represents what may be achievable for quantum information applications, which may require the absorption of multiple simultaneous photons, while Fig. \ref{fig:15502} is closer to the resolution for photon-starved applications which will not see many overlapping photon pulses.

\begin{figure}
\centering\includegraphics[width=3.4in]{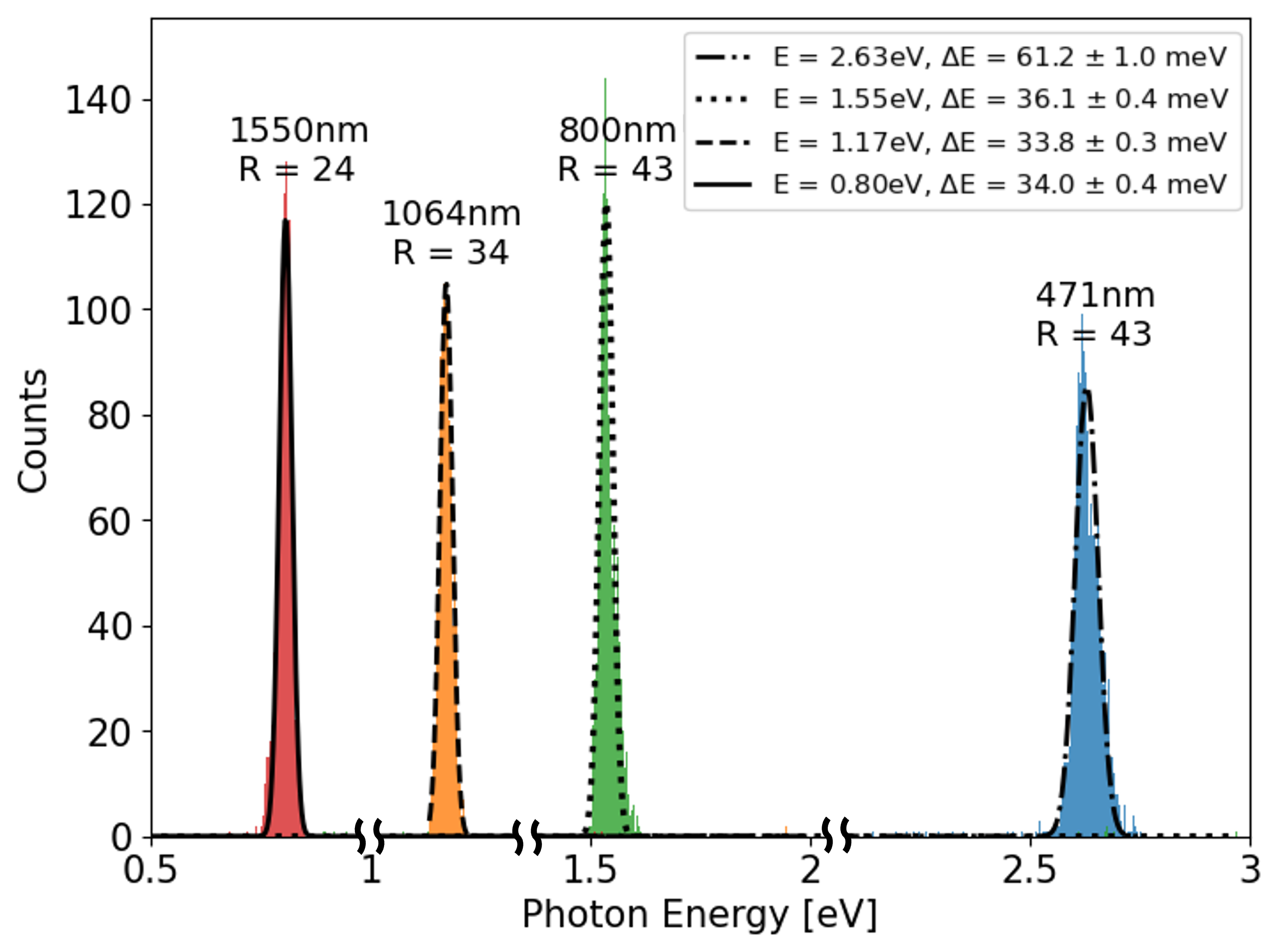}
\caption{\label{fig:multi_pulse}Single-photon peaks for the high resolution TES suspended on a thin membrane with micro-cutouts (Fig. \ref{fig:epsart}a). Single-photon pulse data at 471 nm, 800 nm, 1064 nm, and 1550 nm was collated from four separate measurements at each wavelength, with the $E$ v.s. $\%R_n$ gain separately calibrated for each measurement based on the multi-photon peak heights in the same manner as Fig. \ref{fig:15501}. Each peak represents $\sim1500$ single-photon events incident on the TES, with fitted Gaussian functions used to determine the FWHM energy resolution.}
\end{figure}

\begin{figure}
\centering\includegraphics[width=3.0in]{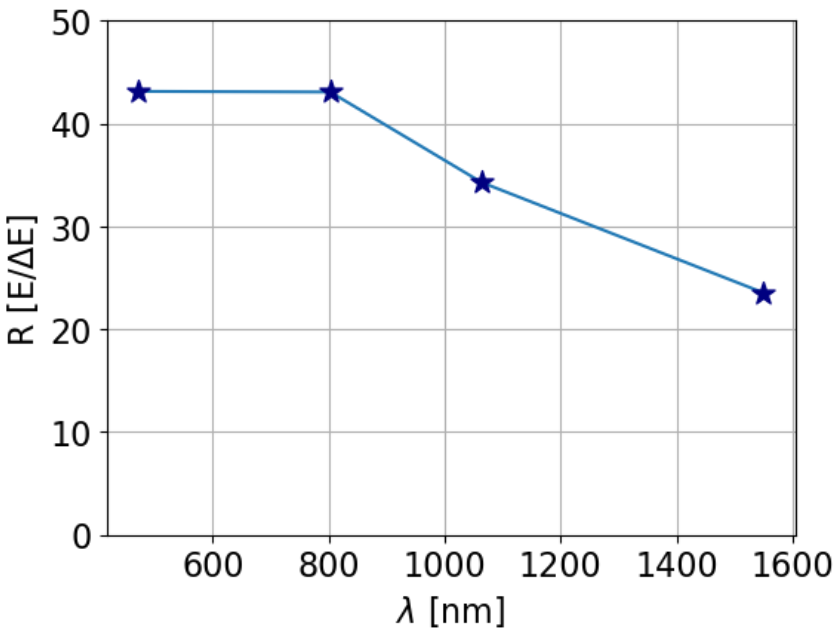}
\caption{\label{fig:R_vs_lam} FWHM energy resolution from Fig. \ref{fig:multi_pulse}, expressed in terms of $R = E/\Delta E$. R values of 43, 43, 34, and 24 were achieved at the measured wavelengths of 471 nm, 800 nm, 1064 nm, and 1550 nm respectively. }
\end{figure}

To determine the broadband resolution of our best performing device design (TES MC), 
we illuminate the device with highly attenuated pulsed laser light at 1550 nm, 1064 nm, 800 nm, and 471 nm wavelengths in three separate measurements. Fig. \ref{fig:multi_pulse} shows the single photon peaks for the four pulsed laser sources used. Each peak represents $\sim1500$ photons, with the overall broadband resolution displayed in Fig. \ref{fig:R_vs_lam}.  Gaussian fits were used to determine the FWHM
energy resolution such as: 34 meV at 0.8 eV (1550 nm photons) corresponding to resolving power $R =24$; 34 meV at 1.17 eV (1064 nm photons) corresponding to $R = 34$; 36 meV at 1.55 eV (800 nm photons) corresponding to $R = 43$ and finally 61 meV at 2.63 eV (471 nm photons) corresponding to an $R=43$.

The broader FWHM of the 471 nm peak (and to a lesser extent the 800 nm peak) and apparent R saturation in Fig. 6 is due to the photon energy exceeding the linear regime of the TES. At 26$\%$ $R_n$, the predicted saturation energy $E_{\textrm{max}} = CT_\textrm{c}(1+\beta)/ \alpha$ is approximately $1.11$ eV. Photons with greater energy than $E_{\textrm{max}}$ will be measured with degraded resolution, since the detector response will become non-linear. However, the TES is still able to detect photons until it is driven completely normal ($100\% \; R_n$), which does not occur until at least $\sim3$ eV by close inspection of Fig. \ref{fig:15501}a-b. Dynamic range optimization of an optical TES would drastically improve the energy resolution at 471 nm, since the current devices were optimized for 1550 nm wavelength, and are outside the linear operation regime for the higher energy photons corresponding to 471 nm wavelength.

This broadband resolution response 
is a strong improvement over prior results for optical TES detectors \cite{Hattori2022, Cabrera1998}, and is also comparable with state-of-the-art optical MKID devices \cite{PhysRevApplied.16.034051}. 

\subsection{\label{sec:phonon}Athermal phonon escape noise}

The main difference between the device configurations of TES MC (on membrane with micro-cutouts, Fig. \ref{fig:epsart}a), TES M (on membrane without cutouts, Fig. \ref{fig:epsart}b), and TES S (on SiN$_\textrm{x}$ layer over substrate, Fig. \ref{fig:epsart}c) is the degree to which the active detector area is isolated from the silicon substrate. The measured energy resolution discrepancy between these device schemes: 
34 meV, 40 meV, and 54 meV respectively, is likely a result of enhanced athermal phonon trapping during the energy down-conversion cascade following absorption of a photon \cite{Kozorezov2006,PhysRevApplied.16.034051}. This is best demonstrated by the $\Delta E_{\textrm{full trapping}}$ row of Table \ref{tab:params}, where the energy resolution values based on Irwin/Hilton model \cite{irwin2005}, as determined by device parameters alone in the absence of athermal phonon escape considerations, are nearly identical for all three device configurations.  

\begin{figure}
\centering\includegraphics[width=3.2in]{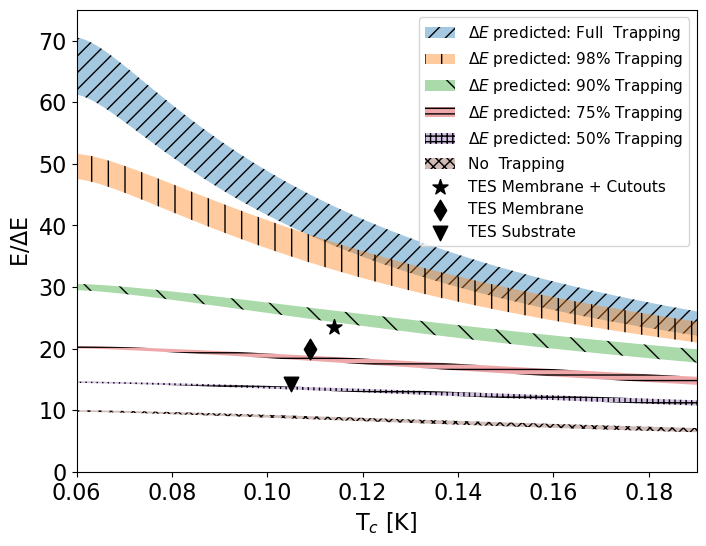}
\caption{\label{fig:ebands} Energy resolution model for 1550 nm light (800 meV), varying the energy collection efficiency $\kappa$, as well as the phonon-escape noise factor $J(E)$, allows us to estimate how close to the ideal Irwin/Hilton model (top blue band) we can feasibly get using phonon trapping methods. At 0\% trapping $\kappa = 0.36$ and $J(E) = 1.22\times10^3$ (unitless). At 100\% trapping, $\kappa = 1$ and $J(E) = 0$, representing zero energy loss to the substrate in the form of athermal phonons. The bands at intermediate degrees of trapping are created by increasing $\kappa$ and reducing $J(E)$ by the specified percentage, a natural method given that both parameters are directly proportional to the amount of energy captured by the TES as thermal energy. Included are three data points corresponding to the measured resolution of the devices of Fig. \ref{fig:epsart}. The models are plotted as bands in order to reflect the variation in TES parameters between the three devices.}
\end{figure}

Modeling the TES energy resolution as a hybrid \cite{ltd_2026} between the athermal phonon escape model described by Kozorezev \cite{Kozorezov2006} and the more idealized model given by Irwin/Hilton \cite{irwin2005}, we can gain insight into the degree of phonon trapping achieved by increasingly isolating the device from the substrate. Fig. \ref{fig:ebands} simulates the degree of phonon trapping using the combined Irwin/Hilton \& Kozorezov model as described in \cite{ltd_2026} and by the following equation:

\begin{equation}
\label{eqn:e_res3}
\text{$\Delta E_{\textrm{predicted}} \approx \;$}
\end{equation}
\begin{equation*}
\text{$ \; 2.355 \sqrt{\frac{4k_\textrm{B} T_\textrm{c} E_{\textrm{max}}}{1+\beta} \sqrt{\frac{nF(T_\textrm{c},T_\textrm{b})}{1-(T_\textrm{b}/T_\textrm{c})^n}}\frac{\sqrt{\xi}}{\kappa^2}+J(E)E},$}
\end{equation*}

\noindent where $T_\textrm{c}$ is the TES superconducting transition temperature, $T_\textrm{b}$ is the bath temperature, and $\beta = (I/R)(dR/dI)$ is the sensitivity of the TES resistance to current. In addition, $n$ is the temperature exponent of power flow to the bath, $F(T_\textrm{c},T_\textrm{b})$ is a unitless function that describes the relative contribution of the bath to the thermodynamics of the TES, and $\xi$ describes a model for the mixed down Johnson noise \cite{Wessels2021}. $E_{\textrm{max}}$ is the maximum photon energy to keep the device in the linear regime, $E_{\textrm{max}} = CT_\textrm{c}(1+\beta)/ \alpha$,  where $C$ is the TES heat capacity and $\alpha = (T/R)(dR/dT)$ is the sensitivity of the TES resistance to temperature. 

The most noteworthy parameters for this discussion are the energy collection efficiency $\kappa$, as well as the statistical variation in phonon loss $J(E)$. For a plain W TES directly on silicon substrate, the stochastic energy loss due to athermal phonons is taken to be  $\kappa = 0.36$ and $J(E) =1.22\times10^3$, which is in-line with literature values \cite{Kozorezov2006}. Varying the factors $\kappa$ and $J(E)$ ($\kappa \rightarrow 1$ and $J(E) \rightarrow 0$) in equation (1), which represents capturing a greater fraction of the incident photon as thermal energy in the electronic system of the device, brings us from the more pessimistic Kozorezov model (bottom brown band in Fig. \ref{fig:ebands}) with no phonon trapping, to the more idealized Irwin/Hilton model (top blue band in Fig. \ref{fig:ebands}) of full phonon trapping. The different degrees of trapping are represented as bands in order to capture the variation of device parameters between the three device schemes listed in Table \ref{tab:params}  (e.g. $\alpha = 205 - 237, \; \beta = 1.18-1.73$).

Plotting the measured $E/\Delta E$ resolving power of the TES MC, TES M, and TES S device configurations on the energy resolution model curves of Fig. \ref{fig:ebands} demonstrates a trend of higher predicted phonon trapping with greater isolation of the TES from the substrate. The lowest resolution device scheme (TES S) displays resolution consistent with 50\% trapping, while the highest resolution device scheme (TES MC) has a measured resolution close to the 90\% trapping band. The increased resolution of the device scheme that is not suspended on membranes (TES S), compared to previous literature, is likely a result of acoustic mismatch between the W TES film and the SiN$_\textrm{x}$ substrate \cite{1979JLTP...37..343K}. Prior measurements showed that our W TES devices on bare Si substrate are consistent with the no trapping band \cite{ltd_2026}. The trend of increased TES superconducting transition temperature ($T_\textrm{c}$) with increased device isolation via membranes and cutouts is thought to be a result of reduced film stress, which has been shown to raise the device $T_\textrm{c}$ \cite{Lita2005}.

\section{Discussion}

The results presented here represent a new energy-resolution record for optical TES detectors, and are comparable with the best results from energy-resolving Microwave Kinetic Inductance Detectors (MKIDs)\cite{PhysRevApplied.16.034051,Zobrist:2022gcj}. The previous best for energy-resolving TESs was obtained by Hattori et al. in 2022 \cite{Hattori2022}, demonstrating $R \leq 12$ at 1550 nm. The record was set by fabrication of a low $T_c$ ($\sim$ 115mK) Au/Ti TES with a small heat capacity ($8\;\upmu$m $\times$ $8\;\upmu$m device area). It is notable that fabrication of our devices on thin membranes has allowed for a higher energy resolution, without requiring a reduction of the active device area ($20\;\upmu$m $\times$ $18\;\upmu$m). This is significant since reduction of the device area can reduce both dynamic range and coupling efficiency, subsequently lowering the system detection efficiency. Moreover, capturing a higher fraction of the photon energy, while essential for enhanced resolution, will effectively reduce the dynamic range. With careful consideration of these limitations, scaling down the device size and lowering the $T_c$ remain viable means to improve the energy resolution of our VIS-NIR TESs even further in future works.

A detailed investigation of device parameters (Table \ref{tab:params}) indicates that, in the absence of phonon-related resolution degradation, the predicted resolution ($\Delta E_{\textrm{full trapping}}$) is nearly equal for all three device schemes. This indicates that phonon loss is likely the primary noise source for our W TES devices. 
Our best performing device (TES MC) measured a record energy resolution of 34 meV at 0.8 eV. This device had a $T_c$ of 114 mK and was fabricated on a thin 75 nm SiN$_\textrm{x}$ membrane with cut-outs in the membrane next to the TES for additional phonon blocking. Based on our modeling displayed in Fig. 7 this device might trap phonons with 90 \%  efficiency, which is a significant improvement compared to the device on solid SiN$_\textrm{x}$ (TES S), which has an efficiency of only 50 \% phonon trapping, and a measured energy resolution of 54 meV at 0.8 eV. The differences between measured energy resolution in Table 1, $\Delta E_{\textrm{meas}}$, and calculated energy resolution based on noise measurements, $\Delta E_{\textrm{noise}}$, indicate a much smaller discrepancy between these values for the devices suspended on membranes (TES MC $\&$ TES M), compared to the device on SiN$_\textrm{x}$ without backside etching of the substrate (TES S). As the device becomes increasingly isolated from the phonon system of the substrate, it appears that the theoretical and measured $\Delta E$ are in better agreement.

To check how well the extracted parameters can model the TES response, and hence the extracted $\Delta E_{\textrm{noise}}$ and $\Delta E_{\textrm{full trapping}}$ values,  we compared the measured time constant of the TES devices with those calculated from the parameters extracted from the complex impedance.
  As seen in Table 1, the predicted time constant based on complex impedance measurements for the TES MC, TES M, and TES S devices deviate from the measured time constant by 4$\%$, 4$\%$, and 11$\%$ respectively, a good indicator for the accuracy of the device parameters extracted from complex impedance measurements and extracted energy resolution values.

Recent studies on dark matter searches \cite{Schwemmbauer} have demonstrated the potential of the optical TES detector as both a target and sensor for direct light dark matter searches due to its intrinsic energy resolution.  The significantly improved energy resolution of TESs on thin membranes, and the possible reduction in background rate due to isolation of detector from possible cosmic events captured in the Si wafer, opens new frontiers for use of optical TESs as sensors for dark matter searches.
In addition to quantum information applications, high energy resolution in the VIS-NIR regime is important for fields of scientific inquiry including exoplanet biosignatures detection (i.e. the Habitable Worlds Observatory), confocal fluorescence laser microscopy \cite{Fukuda2018}, and echelle spectroscopy without cross-dispersion \cite{momos}. The results presented here are a significant step towards enabling these bleeding-edge applications of single-photon detection.

\section{Conclusion}

Strategies for mitigation of athermal phonon loss, which have previously proved effective for energy-resolving optical MKID devices \cite{PhysRevApplied.16.034051}, were also very effective at improving the energy resolution ($\Delta E_{FWHM}$) and resolving power ($E/\Delta E_{FWHM}$) of our tungsten TES devices. By fabricating our devices on very thin (75 nm) membranes, with the addition of micro-cutouts directly adjacent to the photosensitive area, we have achieved a high broadband resolution of 24 - 43 for a band spanning 1550 nm - 471 nm. This result represents a new record among published results for TESs operating in the visible to near infra-red regime.

Through characterization of thermal/electrical device parameters, we were able to estimate the theoretical noise-limited energy resolution of our optical TESs. We found that isolation of the TESs on thin membranes with micro-cutouts brings the resolution closer to theoretical predictions than prior devices fabricated on bare Si substrate \cite{Gerrits2016,ltd_2026}. A combined model of noise-limited resolution, and phonon-loss limited resolution, estimates that our best performing device architecture may achieve an athermal phonon trapping factor close to $90\%$.  Future work aims to push TES energy resolution closer to theoretical predictions and full phonon trapping by using thinner membranes as well as TES designs with lower $T_c$ and different aspect ratio meant to increase critical current and energy resolution \cite{10.1063/1.4984065}.

Combined with the already proven TES characteristics of high system detection efficiency, broad spectrum operability, and near-zero dark counts, our energy resolving W TESs are very well suited to the photon-starved science case of exoplanet biosignatures detection. To properly characterize exoplanet biosignatures, the detector technology will require a resolving power of $R > 100$ at 760~nm (for resolving the O$_2$ biosignature feature, which we believe can be achieved with our further optimized W TESs), and wide spectral coverage of roughly 400~nm to 2000~nm~\cite{rauscher2015, rauscher2018, Nagler2021}. The VIS-NIR TESs presented here have demonstrated robust energy resolution spanning nearly the entire Habitable Worlds Observatory target band. This progress, occurring in parallel with advancements in highly scalable VIS-NIR TES readout schemes \cite{Szypryt2024}, is a very significant step towards the ambitious goal of detecting life on alien worlds.


\begin{acknowledgments}
This work was supported in part by the NASA APRA program under Grant No. NNH23OB118A  and Xanadu Quantum Technologies (https://ror.org/05qj0n416).

\end{acknowledgments}

\bibliography{refs}

\appendix{Supplemental}

\maketitle

\section{Device characterization}

We model our W TES devices under the usual conditions for calorimetric operation, strong electrothermal
feedback with stiff voltage bias in the small-signal limit.
We evaluated the superconducting transition characteristics of our W TES devices by measuring the I–V curves at various bath temperatures $T_{\textrm{bath}}$. At each temperature, the Joule power P was extracted at
the bias point where the TES resistance equals 0.8 normal resistance $R_n$. 
We assume a power-law dependence of thermal flow to the heat bath given by \cite{irwin2005}:

\begin{equation}
    P = k(T^n - T_{\textrm{bath}}^n),
\end{equation}

\noindent with thermal conductance exponent $n$, and constant prefactor $k = G/nT^{n-1}$ associated with the thermal conductance by $G=\textrm{d}P/\textrm{d}T$.
The TES parameters extracted from the IV curves are listed in Table \ref{tab:params} for each of the three devices in Figure 1 of the main document.

Subsequently, we measure the device heat capacity ($C$) by raising the cryostat base temperature until the device is extremely high in transition ($T_{\textrm{bath}} \rightarrow T_{\textrm{c}}$) so that the effects of electrothermal feedback become negligible. In this regime, the slow exponential decay of TES temperature after absorption of a single photon laser pulse (1550 nm in this case) may be fit to extract the heat capacity from the intrinsic thermal decay time constant: $\tau = {C}/{G}$.







Since the photon energy absorbed in the TES is simultaneously transduced into the thermal signal (temperature increase) and electrical signal (resistance increase), the thermal ($\alpha$) and electrical ($\beta$) sensitivities of the TES ($R(I,T)$) are defined as:

\begin{equation}
    \alpha = \left. \frac{T_0}{R_0}\frac{\partial R}{\partial T} \right|_{I_0}, \; \; 
    \beta = \left. \frac{I_0}{R_0}\frac{\partial R}{\partial I} \right|_{T_0},
\end{equation}

\noindent where $R_{0}$, $I_0$, and $T_0$ represent the instantaneous steady-state values of resistance, current, and temperature respectively. The parameters $R$, $\alpha$ and $\beta$ are commonly used to describe the behavior of $R(I,T)$ in the small signal limit \cite{irwin2005}. We also define the loop gain under constant current:

\begin{figure*}
\centering\includegraphics[width=5.0in]{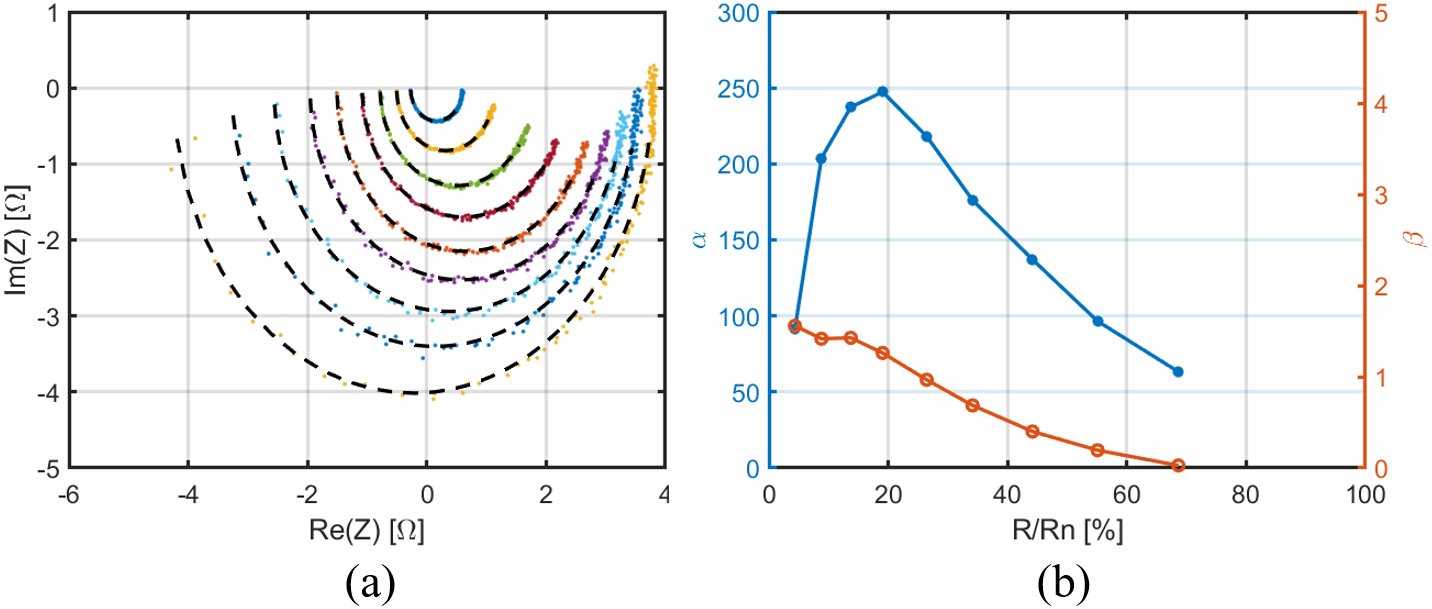}
\caption{\label{fig:complexz }(a) Complex impedance data for the device scheme on the SiNx membrane with micro-cutouts. Complex impedance data was fitted as in \cite{Lindeman} in order to extract the loop-gain $\mathscr{L}$ and subsequently extract $\alpha$, $\beta$, and $\tau^-_{\textrm{calc}}$ (Table \ref{tab:params}). (b) $\alpha$ and $\beta$ plotted as a function of \%$R_\textrm{n}
$. Notably, both $\alpha$ and $\beta$ peak early on in the transition, which is also where we find the highest $E/\Delta E$. }
\end{figure*}

\begin{equation}
\label{eqn:gain}
    \mathscr{L} = P_{J0}\alpha/GT_0 ,
\end{equation}

\noindent where $P_{J_0}$ is the instantaneous steady-state joule power ($I_0^2R_0$) dissipation.

In a manner consistent with Irwin's chapter of ``Cryogenic Particle Detection" \cite{irwin2005}, we model the power to current responsivity in response to a time varying electrical signal with angular frequency $\omega$, computed by solving the coupled thermodynamic/electrical equations governing the TES, as:

\begin{equation}
    \text{$s_I(\omega) = -\left(  I_0R_0 \left[ \frac{L}{\tau_{el}R_0\mathscr{L}} +\left( 1-\frac{R_L}{R_0} \right) +  \right. \right. $}\\
\end{equation}
\begin{equation*}
    \text{$ \left. \left. \textrm{i}\omega\frac{L\tau}{R_0\mathscr{L}} \left( \frac{1}{\tau_I} + \frac{1}{\tau_{el}} \right) - \frac{\omega^2\tau L}{\mathscr{L} R_0}  \right]  \right)^{-1}$,}
\end{equation*}

\noindent where $L$, $\tau_I$, and $\tau_{el}$ are the input inductance measured from complex impedance measurements, the time constant of the TES under stiff voltage bias, and the time constant of the electrical bias circuit, respectively. The TES responsivity, $s_I(\omega)$, allows us to estimate all the noise components for our TES devices and ultimately its fundamental noise-limited energy resolution, which will be demonstrated in Section \ref{sec:noise}.

\subsection{Complex Impedance Measurements}

 
 We used measurements of TES complex impedance as a function of frequency to extract the thermal ($\alpha$) and electrical ($\beta$) sensitivities characterizing the TES resistance and calculate the values for current noise and expected energy resolution.  We subsequently compare to measured values for current noise and energy resolution.
 The complex impedance of a single-block thermal model TES at an angular frequency $\omega$ can be expressed as follows  \cite{irwin2005}:

\begin{equation}
\label{eq:comz}
    Z = R_0(1+\beta_I)+\frac{R_0\mathscr{L}_I(2+\beta_I)}{(1-\mathscr{L}_I)(1+\textrm{i}\omega\tau_I)} \;.
\end{equation}

We measured complex impedance by injecting a small AC signal (22 nV across the TES) from a network analyzer into the voltage-biased TES. The complex impedance data can be well described using Equation \ref{eq:comz}. Figure \ref{fig:complexz }a shows the complex impedance measurement of device TES MC (on membrane with micro-cutouts), with corresponding single-block TES model data fits that overlap nicely. The fits to complex impedance data are first used to determine the thermal sensitivities $\alpha$ and $\beta$ \cite{Lindeman}. Then, the previously determined thermal parameters ($G$, $k$, $n$, $C$), may be combined with $\mathscr{L}$ (Equation \ref{eqn:gain}) to extract the remaining electro-thermal parameters (time constant and $s_I(\omega)$). The values of the thermal/electrical sensitivities ($\alpha$, $\beta$) as a function of TES resistance (bias point on the transition curve displayed as $\%R_n$) are given in Figure \ref{fig:complexz }b.

 To check how well the extracted parameters can model the TES response, we compared the measured time constant of the TES with that calculated from the parameters extracted from the complex impedance. 
When the shunt resistance in the bias circuit is negligible, the fall time $\tau_{eff}$ ($\tau_{calc}$ in  TABLE I) can be expressed as follows \cite{irwin2005}:

\begin{equation}
\label{eq:teff}
    \tau_{eff} = \frac{C}{G} \frac{(1+\beta)}{(1+\beta +\mathscr{L})}
\end{equation}

 The calculated time constants using Equation \ref{eq:teff} are comparable with the measured time constant (deviation of 4$\%$, 4$\%$, and 11$\%$ for the TES MC, TES M, and TES S devices respectively) as shown in TABLE I.


\subsection{\label{sec:noise}Noise-limited energy resolution}

TES detectors are susceptible to thermodynamic noise due to fluctuations in both current (Johnson noise) and temperature (thermal fluctuation noise). In this section, we quantify the effects of all relevant noise sources (except for athermal-phonon escape) by calculating the overall spectral density of the noise.
Once the total current noise (${S_{I_{tot}}}$) is defined, we can calculate the theoretical noise-floor of our TES devices based on \cite{McCammon2005}:

\begin{equation}
\label{eq:delnoise}
    \Delta E_{noise} = 2.355 \left( \int_0^\infty \frac{4|s_I(\omega)|^2}{S_{I_{tot}}}\textrm{d}f\right)^{-1/2} \; .
\end{equation}

To start, we separate ${S_{I_{tot}}}$ into its electronic/thermal and readout noise elements:

\begin{equation}
\label{eq:noisetot}
    S_{I_{tot}} = S_{I_{johnson}} + S_{I_{{L}}} + S_{I_{{TFN}}} + S_{I_{{SQUID}}} \; .
\end{equation}

\noindent where the Johnson noise is expressed, in current spectral density form, as:

\begin{equation}
    S_{I_{johnson}} = 4k_B T_0I_0^2R_0 (1+2\beta_I)(1+\omega^2\tau^2)\frac{|s_I(\omega)|^2}{\mathscr{L}_I^2} \; ,
\end{equation}

\noindent and similarly, the Johnson noise of the shunt and parasitic resistances is given by:

\begin{equation}
    S_{I_{{L}}}= 4k_B T_0I_0^2R_L (1+2\beta_I)(1+\omega^2\tau^2)\frac{|s_I(\omega)|^2}{\mathscr{L}_I^2} \; ,
\end{equation}

\noindent with $R_L = R_{shunt}+R_{par}$.

With the electronic portion of the noise accounted for, we quantify the thermal fluctuation noise as:

\begin{equation}
    S_{I_{{TFN}}}  = 4k_B T_0^2G|s_I(\omega)|^2 \times F(T_0, T_{bath}) \;,
\end{equation}

\noindent where $F(T_0,T_{bath})$ is taken, as in 
\cite{Mather:82}, to be:

\begin{equation}
\label{eq:F}
    F(T_0, T_{bath}) = \left( \left(\frac{T_{bath}}{T_0} \right)^{n+1} +1\right) \bigg/2
\end{equation}

Figure \ref{fig:noise} shows the noise components calculated using Equations \ref{eq:noisetot} - \ref{eq:F} with the TES physical characteristics in Table \ref{tab:params}. All plots are for the TES MC device biased at the point in transition giving the highest measured $E/\Delta E$ (13$\%\;R_n$, near the point of highest $\alpha$). We also plot the measured noise (the average of 200 noise traces spanning $\sim 10^3 -10^5$ Hz) for both the SQUID readout in isolation 
(Readout noise), as well as the entire TES + readout system. We see from Figure \ref{fig:noise} that the sum total of the theoretical noise ($S_{I_{tot}}$, black curve) is very much in line with the experimentally measured noise (red curve). 

Returning to Equation \ref{eq:delnoise}, we may integrate over the experimentally measured, or the theoretically estimated, noise to arrive at a prediction for $\Delta E_{noise}$. The values for all three devices, calculated from the experimental noise measurements, are given in the $\Delta E_{noise}$ row of Table \ref{tab:params}.

\end{document}